\documentclass[aps,pre,showpacs,noshowkeys,amsmath,amssymb,amsfonts,superscriptaddress,longbibliography,reprint]{revtex4-2}
\UseRawInputEncoding
\usepackage[english]{babel}

\usepackage{graphicx}
\usepackage{bm}
\usepackage{physics}
\usepackage{mathtools}
\usepackage{gensymb}

\usepackage{caption}
\usepackage{subcaption}
\DeclareCaptionLabelSeparator{bar}{~\rule[-0.4ex]{0.2ex}{1em}~}
\DeclareCaptionLabelFormat{subfor}{\textbf{#2}}
\newcommand*\bfcaption[2]{\caption[#1]{\textbf{#1.}#2}}
\usepackage[dvipsnames]{xcolor}
\definecolor{UBcolor}{HTML}{007CC1}

\usepackage[colorlinks=true,pdfnewwindow=true,linkcolor=UBcolor,citecolor=UBcolor,urlcolor=UBcolor,breaklinks=true,linktocpage]{hyperref}
\usepackage[all]{hypcap}
\usepackage[nameinlink,capitalise]{cleveref}
\usepackage{xr}
\crefname{SI section}{SI Section}{SI Sections}
\Crefname{SI section}{SI Section}{SI Sections}
\begin{document}

\title{Bistability of cellular traction on strain-stiffening substrates}

\author{Irina Pi-Jaum\`{a}}
\affiliation{Departament de F\'{i}sica de la Mat\`{e}ria Condensada, Universitat de Barcelona, 08028 Barcelona, Spain}
\affiliation{Universitat de Barcelona Institute of Complex Systems (UBICS), 08028 Barcelona, Spain}

\author{Jaume Casademunt}
\affiliation{Departament de F\'{i}sica de la Mat\`{e}ria Condensada, Universitat de Barcelona, 08028 Barcelona, Spain}
\affiliation{Universitat de Barcelona Institute of Complex Systems (UBICS), 08028 Barcelona, Spain}

\author{Ricard Alert}
\email{ricard.alert@ub.edu}
\affiliation{Departament de F\'{i}sica de la Mat\`{e}ria Condensada, Universitat de Barcelona, 08028 Barcelona, Spain}
\affiliation{Universitat de Barcelona Institute of Complex Systems (UBICS), 08028 Barcelona, Spain}
\affiliation{Instituci\'{o} Catalana de Recerca i Estudis Avan\c{c}ats (ICREA), Barcelona, Spain}
\affiliation{Max Planck Institute for the Physics of Complex Systems, 01187 Dresden, Germany}
\affiliation{Center for Systems Biology Dresden, 01307 Dresden, Germany}
\affiliation{Cluster of Excellence Physics of Life, TU Dresden, 01062 Dresden, Germany}

\begin{abstract}
To migrate, cells exert traction forces on the extracellular matrix (ECM) --- a biopolymer network that often exhibits nonlinear strain-stiffening elasticity. Cellular tractions can therefore stiffen the ECM. At the same time, cells exert stronger tractions on stiffer ECM. Here, we show theoretically that this traction-stiffness feedback can produce traction bistability and hysteresis. As a result, increasing either the ECM's nonlinear elasticity or cellular contractility leads to a discontinuous transition from low to high tractions. This traction jump might trigger collective cell migration as the ECM stiffens, for example during development and tumor progression. Moreover, the bistable behavior might provide robustness to cellular traction forces when cells migrate through mechanically heterogeneous environments.
\end{abstract}

\maketitle 

Cells are mechanically active. They exert traction forces on the surrounding extracellular matrix (ECM), which allows them to migrate and to probe their environment. The environment, however, is not merely a passive scaffold. The ECM is a biopolymer fiber network that cells secrete, remodel, and degrade \cite{Pally2024}. For instance, fibroblasts secrete fibronectin and collagen fibers around wounds and tumors, which stiffens the tissue \cite{Barbazan2023}. Cancer cells can degrade the matrix using metalloprotease enzymes \cite{Nabeshima2000,Wolf2007}, and can also align the ECM fibers by pulling on them, which promotes tumor invasion \cite{Friedl2003,Friedl2011,Winkler2020,Ray2021,Provenzano2006,Kopanska2016,Glentis2017,Li2017,Ahmadzadeh2017,Meng2024,Kim2026}. The remodeling of the ECM by cell migration is also instrumental for morphogenesis, for example to ensure the elongation of \textit{Drosophila} eggs \cite{Haigo2011}. Such cell-ECM interactions are mostly bidirectional --- an observation that has been named \emph{mechanoreciprocity} \cite{VanHelvert2018}, and the many consequences of this interplay are under intense investigation \cite{Pally2024,Adar2021,Adar2022,Adar2024,Palmquist2022,Clark2022,Chepizhko2025,Zhang2025,Gottheil2024}.

\begin{figure}[tb!]
    \centering
    {{\includegraphics[width=\columnwidth]{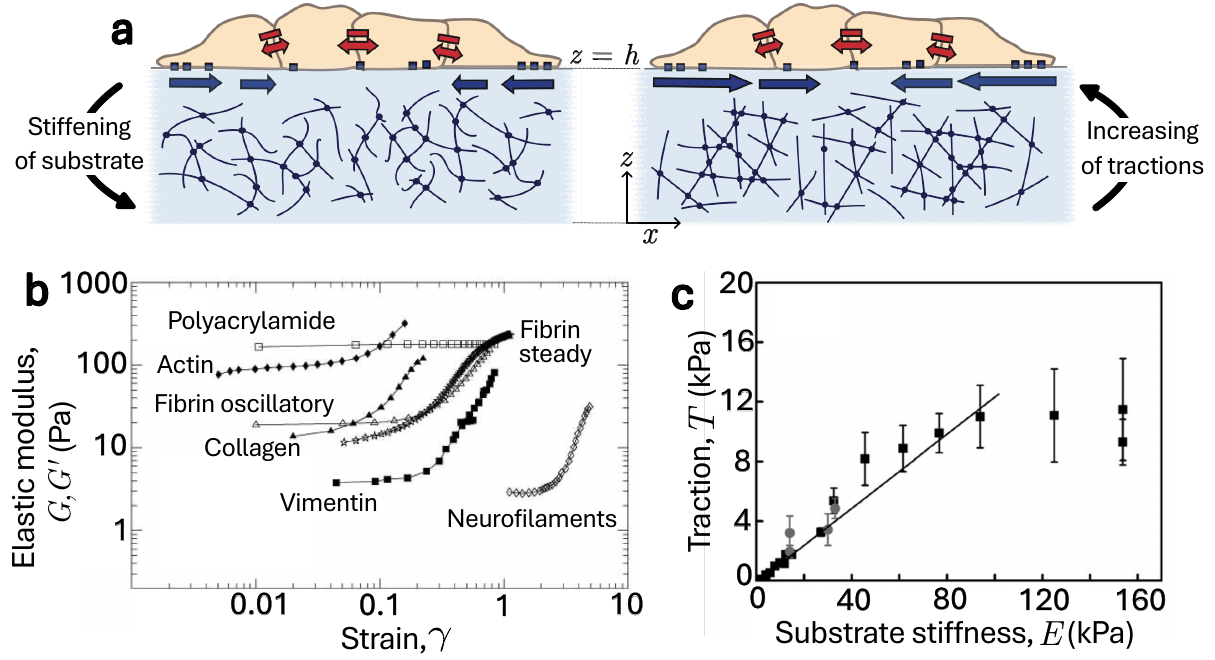}}}
  {\phantomsubcaption\label{Fig feedback}}
  {\phantomsubcaption\label{Fig strain-stiffening}}
  {\phantomsubcaption\label{Fig traction-stiffness}}
    \bfcaption{Feedback between cellular tractions and substrate stiffness}{ \subref*{Fig feedback}, Traction forces (blue arrows) can stiffen the substrate. Stiffer substrates (represented via tense fibers) lead to stronger tractions. \subref*{Fig strain-stiffening}, Experimental measurements of strain stiffening for diverse biopolymer networks. Adapted from Ref. \cite{Storm2005}. \subref*{Fig traction-stiffness}, Experimental measurements of traction forces increasing with substrate stiffness. Adapted from Ref. \cite{Ghibaudo2008}.}
    \label{fig:sketch}
\end{figure}

Here, we address this issue by focusing on the feedback between cellular tractions and ECM stiffness (\cref{Fig feedback}). On the one hand, cellular tractions deform the ECM. In fact, these deformations can be used to infer the tractions that caused them, which is the principle behind the technique called traction force microscopy (TFM) \cite{Trepat2009,Gomez-Gonzalez2020}. Measurements using TFM are mostly performed in vitro by placing cells on a synthetic hydrogel substrate, made for example of polyacrylamide, chosen for its wide range of linear elastic response. However, most physiological ECM networks --- consisting for example of collagen, fibronectin, and fibrin --- exhibit a marked nonlinear elastic behavior. In particular, they display strain stiffening: Their elastic shear modulus increases strongly above a certain strain (\cref{Fig strain-stiffening}) \cite{Broedersz2014,Alisafaei2021,Storm2005,Licup2015,Hall2016,Han2018,Song2025,EnriquezMartinez2025}. Epithelial cell layers, which can also act as substrates for migration, also display strain stiffening \cite{Duque2024}. This stiffening is likely physiologically relevant as a means to prevent large deformations that could damage tissues \cite{Storm2005}.

On the other hand, cells sense and respond to the ECM stiffness. Several studies over the past two decades have shown that most cells exert stronger tractions on stiffer substrates (\cref{Fig traction-stiffness}) \cite{Discher2005,Ladoux2012,Gupta2016,Janmey2020,Saez2005,Ghibaudo2008,Saez2010,Ladoux2010,Trichet2012,Elosegui-Artola2014,Breckenridge2014,Gupta2015}. This cellular response underlies durotaxis --- the migration from softer to stiffer substrates \cite{Breckenridge2014,Alert2019a,Pi-Jauma2022,Pallares2023}. Thus, by deforming the environment, cellular tractions can stiffen it \cite{Hall2016,Han2018,Yang2023,Alisafaei2021}, which, in turn, leads to stronger tractions \cite{Hall2016,EnriquezMartinez2025}.

We propose a minimal model of this positive feedback between cellular tractions and substrate stiffness. We find that, as a result of this feedback, the traction magnitude undergoes a transition between low and high values. We determine the conditions for this transition in terms of the two exponents characterizing the nonlinear effects of the model: the increase of tractions with stiffness, and matrix strain stiffening. We then discuss two ways in which this transition could take place in experiments: (i) by increasing the nonlinear elastic response of the substrate, for example by increasing the ECM polymer concentration, and (ii) by increasing cellular force generation, for example by phosphorylating myosin molecular motors. Our findings provide a mechanism for cells to switch from a low- to a high-traction state, which might trigger collective cell migration in processes such as embryo implantation \cite{Godeau2025,Cavanaugh2025}, development \cite{Barriga2018,Barriga2019}, and tumor invasion \cite{Mierke2008,Levental2009,Koch2012,Kraning-Rush2012,Krajina2021,Meng2024}.

\vspace{0.4cm}
\textbf{Model for traction-stiffness feedback.} To illustrate the feedback between tractions and substrate stiffness in a simple setting, we consider a two-dimensional section of an elastic substrate, infinite along the $\hat{\mathbf{x}}$ axis and with a height $h$ along the $\hat{\mathbf{z}}$ axis, with a one-dimensional cell train on top (\cref{Fig feedback}). The cell train can be of any length; it can represent from a single cell to a section of a cell monolayer. Following the theory that underlies TFM \cite{Trepat2009}, we consider the strain that the traction forces generate in the substrate. We treat the substrate as an elastic material with shear modulus $G$ and bulk modulus $K$, which follows the constitutive relation
\begin{align}
    \sigma_{ij} &= 2G(\gamma_{xz}) \left( \gamma_{ij} - \frac{1}{d}\gamma_{kk}\delta_{ij}\right) + K\gamma_{kk}\delta_{ij} \label{eq:stress}
\end{align}
between the stress tensor $\sigma_{ij}$ and the strain tensor $\gamma_{ij} = \frac{1}{2}\left(\partial_i u_j + \partial_j u_i \right)$, with $u_i$ being the elastic displacement, and $d$ the dimensionality.

While the theory of TFM considers a linearly elastic substrate \cite{Trepat2009}, here we capture the strain-stiffnening behavior characteristic of biopolymer networks by making the shear modulus $G$ a function of the shear strain $\gamma_{xz}$. To account for this nonlinear response in a simple way, we assume that the linear response of the material, with a constant shear modulus $G_0$, extends up to a critical strain $\gamma_\text{c}$, above which there is a nonlinear elastic response given by a power law with exponent $\alpha$:
\begin{equation}  \label{eq:G(gamma)}
    G(\gamma_{xz}) =  \begin{cases} 
      G_0, & \gamma_{xz} \leq \gamma_\text{c}, \\
      G_1 \gamma_{xz}^\alpha, & \gamma_{xz} >  \gamma_\text{c}.
   \end{cases}
\end{equation}
Here, we call $G_1$ the nonlinear elastic coefficient, which satisfies $G_0 = G_1 \gamma_\text{c}^\alpha$ to ensure continuity of the elastic response.

Force balance in the substrate is given by $\bm{\nabla}\cdot\bm{\sigma} = 0$, which is written in components as
\begin{subequations} \label{eq:force_balance}
\begin{align}
      \partial_x \sigma_{xx} + \partial_z \sigma_{xz} &= 0, \\ 
      \partial_x \sigma_{zx} + \partial_z \sigma_{zz} &= 0.
\end{align}
\end{subequations}
At $z=0$, the substrate gel is attached to the bottom glass dish, which imposes no-displacement conditions. Respectively, at $z=h$, we consider that the normal stress is negligible, and that the shear stress is given by the traction that the cells exert on the substrate. Thus, we have
\begin{subequations} \label{eq:BC}
\begin{align}
    u_x\big|_{z=0} &= 0, \quad u_z\big|_{z=0} = 0 \label{eq:BC_bottom} \\
    \sigma_{zz}\big|_{z=h} &= 0,  \quad \sigma_{xz}\big|_{z=h} = T_x \equiv T. \label{eq:BC_top}
\end{align}
\end{subequations}

\begin{figure}[tb!]
    \centering
    {{\includegraphics[width=\columnwidth]{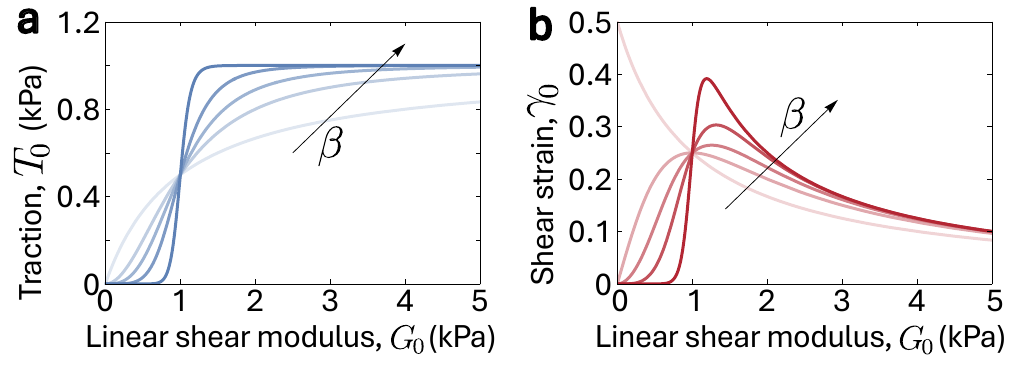}}}
  {\phantomsubcaption\label{Fig traction-linear}}
  {\phantomsubcaption\label{Fig strain-linear}}
    \bfcaption{Stationary states in the linear regime}{ Traction $T_0$ (\subref*{Fig traction-linear}) and substrate shear strain $\gamma_0$ (\subref*{Fig strain-linear}) as a function of the linear shear modulus $G_0$ of the substrate. The plots represent \cref{eq:Ttarget_Hill} with $G=G_0$, and \cref{eq:stat_linear_Hill}, with $T_\infty = 1$ kPa, $G_* = 1$ kPa, and for values of $\beta = 1,2,3,5,15$.}
    \label{fig:Linear_Hill}
\end{figure}

Finally, we assume that cells adapt their traction in response to substrate stiffness over a time scale $\tau$, which reflects the duration of mechanosensing and mechanotransduction processes, as well as the time it takes for cells to remodel cell-substrate adhesions. Thus, we propose the relaxation equation
\begin{equation}  \label{eq:dynamics}
    \partial_t T = -\frac{1}{\tau} \left[ T-T_\text{target} \left(G \big|_{z=h} \right) \right],
\end{equation}
where $T_\text{target}$ is the traction value that the cells reach upon adaptation. This target traction is a function of the shear modulus $G|_{z=h}$ at the substrate's upper surface. Cellular tractions often increase and eventually saturate with increasing substrate stiffness \cite{Ghibaudo2008,Ladoux2010,Trichet2012,Gupta2015}. To capture this cellular response, we consider a Hill function of exponent $\beta$:
\begin{equation} 
    T_{\text{target}}(G) =  T_\infty \frac{G^\beta}{G^\beta+G_*^\beta}, \label{eq:Ttarget_Hill} 
\end{equation}
where $T_\infty$ is the maximal traction at saturation, and $G_*$ is the characteristic shear modulus above which traction saturates (\cref{Fig traction-linear}). Previous studies have predicted \cite{Zemel2010,Walcott2010,Marcq2011,Sens2013,Gupta2015} and used \cite{Alert2019a,Pi-Jauma2022,Pallares2023} a Hill function of order 1. However, as recently observed experimentally in muscle tissue \cite{Kah2025}, the cellular response could be sharper, represented here by higher values of the exponent $\beta$ (\cref{Fig traction-linear}).

\vspace{0.4cm}
\textbf{Stationary solutions.} To illustrate the behavior of the model, we consider a situation with uniform traction. In this case, indicated below by a superscript $0$, the solution to the force balance \cref{eq:force_balance} with the boundary conditions in \cref{eq:BC} gives $u_z^0 = 0$ and $u_x^0 = cz$, with $c$ a constant given by $c=T/G_0$ in the linear elastic regime and by $c=2 (T/(2G_1))^{1/(1+\alpha)}$ in the nonlinear regime. Consequently, the only non-zero component of the strain tensor is $\gamma_{xz}^0 = c/2$. Thus, hereafter we drop the indices and denote $\gamma^0_{xz} \equiv \gamma_0$. At the stationary state, the traction is $T_0 = T_\text{target}(G(\gamma_0))$. Thus, using \cref{eq:Ttarget_Hill} and introducing it into the $\sigma_{xz}$ boundary condition in \cref{eq:BC_top}, we obtain the following equation for the stationary shear strain $\gamma_0$:
\begin{align} \label{eq:gamma_stat}
    2G(\gamma_0) \gamma_0 = T_\infty\frac{G(\gamma_0)^\beta}{G(\gamma_0)^\beta + G_*^\beta},
\end{align}
where $G(\gamma_0)$ is given by \cref{eq:G(gamma)}.

\vspace{0.2cm}
\emph{Linear regime ---} We first solve \cref{eq:gamma_stat} in the linear regime ($\gamma_0 \leq \gamma_\text{c}$), for which $G = G_0$ is independent of the strain. The solution then is
\begin{equation}
    \gamma_0 =  \frac{T_\infty}{2} \frac{G_0^{\beta-1}}{G_0^\beta+G_*^\beta}.  \label{eq:stat_linear_Hill} 
\end{equation}
This solution shows that, whereas tractions increase monotonically (\cref{Fig traction-linear}), the substrate strain is non-monotonic with substrate stiffness (\cref{Fig strain-linear}). The reason is that, for $\beta >1$, tractions grow sufficiently strongly with stiffness at small $G_0$ to produce an increasing strain as the stiffness increases. However, at larger stiffness $G_0$, tractions tend to saturate, and hence the strain decreases as the substrate becomes harder to deform. The maximal substrate strain takes place at a substrate shear modulus $G_0^\text{max} = G_* (\beta - 1)^{1/\beta}$.

\begin{figure}[tb!]
    \centering
    {{\includegraphics[width=\columnwidth]{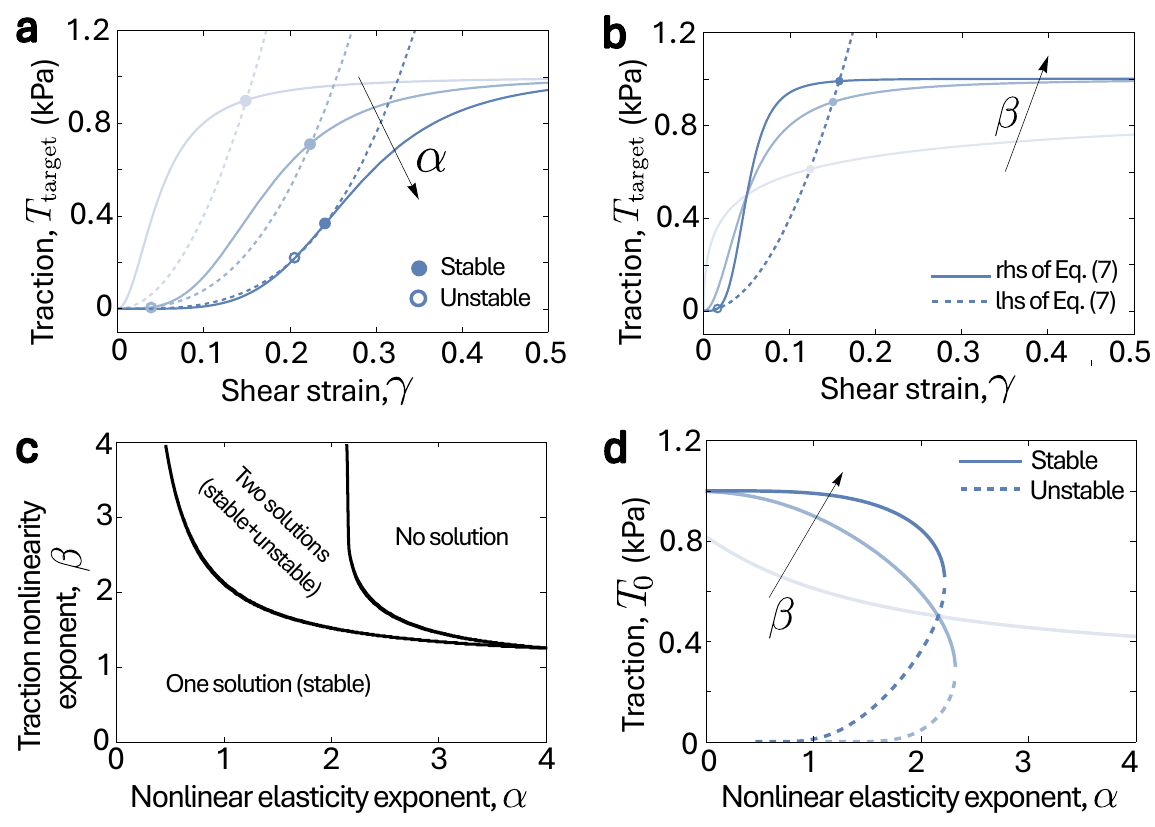}}}
  {\phantomsubcaption\label{Fig equation-nonlinear-alpha}}
  {\phantomsubcaption\label{Fig equation-nonlinear-beta}}
  {\phantomsubcaption\label{Fig phase-diagram-alpha-beta}}
  {\phantomsubcaption\label{Fig bifurcation}}
    \bfcaption{Stationary states in the nonlinear regime}{ \subref*{Fig equation-nonlinear-alpha},\subref*{Fig equation-nonlinear-beta}, Solid and dashed lines are the rhs and the lhs of \cref{eq:gamma_stat}, respectively. Their intersection determines the stable (filled dots) and unstable (empty dots) stationary solutions of traction $T_0$ and shear strain $\gamma_0$. In \subref*{Fig equation-nonlinear-alpha}, $\beta=2$ is fixed and $\alpha = 1,1.7,2.3$ varies. In \subref*{Fig equation-nonlinear-beta}, $\alpha = 1$ is fixed, and $\beta = 0.5,2,4$ varies. \subref*{Fig phase-diagram-alpha-beta}, Diagram of solutions in the nonlinear regime as a function of the nonlinearity exponents $\alpha$ and $\beta$. \subref*{Fig bifurcation}, Bifurcation diagram showing the stable (solid) and unstable (dashed) solutions for the traction as a function of the exponent $\alpha$, for $\beta = 0.5,2,4$. For sufficiently high $\alpha$ and $\beta$, the stable and unstable solutions annhilate in a saddle-node bifurcation. In all panels, $G_1 = 20$ kPa, $G^*=1$ kPa, $T^\infty = 1$ kPa and $\gamma_{\text{c}}=0.2$.}
    \label{fig:Nonlinear_Hill}
\end{figure}

\vspace{0.2cm}
\emph{Nonlinear regime ---} In the nonlinear regime, for which $G=G_1\gamma_0^\alpha$, \cref{eq:gamma_stat} becomes transcendent, and we therefore solve it numerically. The non-vanishing solutions correspond to the dots in \cref{Fig equation-nonlinear-alpha,Fig equation-nonlinear-beta}, which mark the intersections between the left-hand side (dashed lines) and the right-hand-side (solid lines) of \cref{eq:gamma_stat}. When the nonlinearity exponents $\alpha$ and $\beta$ are small, there is a single stationary solution with non-vanishing traction. For larger values of $\alpha$ and $\beta$, there are two non-vanishing stationary solutions (\cref{Fig equation-nonlinear-alpha,Fig equation-nonlinear-beta}), corresponding to states of lower and higher traction. For even larger values of $\alpha$ and $\beta$, there are no solutions (\cref{Fig phase-diagram-alpha-beta}), since the previous stable and unstable solutions annihilate in a saddle-node bifurcation (\cref{Fig bifurcation}).

In the region with two solutions, to determine their stability, we perform a linear stability analysis by considering $\gamma = \gamma_0 + \delta \gamma$ and $T = T_0 + \delta T$. From \cref{eq:dynamics}, the perturbation dynamics reads
\begin{equation}  \label{eq:dynamics_perturb}
    \partial_t \delta T = -\frac{1}{\tau} \left[ \delta T- \left. \frac{\partial  T_{\text{target}}}{\partial \gamma} \right|_{\gamma_0} \delta \gamma \right].
\end{equation}
To eliminate $\delta\gamma$, we use force balance at the cell-substrate surface. Combining the boundary condition in \cref{eq:BC_top}, $\sigma_{xz} = T$, and the substrate's constitutive relation \cref{eq:stress}, $\sigma_{xz} = 2G(\gamma) \gamma$, we have $2G(\gamma) \gamma = T$ for uniform traction. Expanding this condition to first order in perturbations, we obtain
\begin{equation} \label{eq:gammaxz_h}
\delta \gamma = \frac{\delta T}{2 \left[ G(\gamma_0) + G'(\gamma_0) \gamma_0 \right] } = \frac{\delta T}{ 2G_1 \gamma_0^\alpha (1+\alpha)},
\end{equation}
where in the second equality we have introduced the nonlinear $G(\gamma)$ relation from \cref{eq:G(gamma)}. Introducing this result in \cref{eq:dynamics_perturb} and computing the derivative of the target traction that appears in that equation, we finally obtain the growth rate of traction perturbations:
\begin{equation} \label{eq:growth-rate}
\omega = - \frac{1}{\tau} \left[ 1 - T_\infty \frac{\alpha \beta}{2(1 + \alpha)} \frac{ G_*^\beta G_1^{\beta - 1} \gamma_0^{\alpha\beta - \alpha - 1} }{ \left[ (G_1 \gamma_0^\alpha)^\beta + G_*^\beta \right]^2 } \right].
\end{equation}
A given solution, with strain $\gamma_0$ obtained numerically from \cref{eq:gamma_stat}, is stable if $\omega <0$ (filled dots in \cref{Fig equation-nonlinear-alpha,Fig equation-nonlinear-beta} and solid lines in \cref{Fig bifurcation}) and unstable if $\omega>0$ (empty circles in \cref{Fig equation-nonlinear-alpha,Fig equation-nonlinear-beta} and dashed lines in \cref{Fig bifurcation}). 

This analysis shows that, in the regime with two solutions, the low-strain low-traction solution is unstable. At low strain, the substrate is softer, and tractions increase steeply with its stiffness, which promotes the instability. Instead, at high strain, the substrate is stiffer, and tractions tend to a saturate, which stabilizes the high-strain high-traction state. In summary, if tractions increase sufficiently steeply with stiffness and the substrate strain-stiffens strongly enough, i.e. if both $\beta$ and $\alpha$ are sufficiently large, the system will transition from a low-traction to a high-traction stable state (\cref{Fig phase-diagram-alpha-beta}).

Finally, in the SI, we generalize our analysis to account for non-uniform tractions, which lead to depth-dependent strains (\cref{sec_SI:stability}) and are stabilized at short wavelengths by polarity alignment between neighboring cells (\cref{sec_SI:diffusion}).

\vspace{0.4cm}
\textbf{Traction bistability and hysteresis.} We study the behavior of the substrate strain and cellular traction as a function of the nonlinear elasticity coefficient $G_1$ and the saturation traction $T_\infty$, since these parameters could be tuned experimentally. For example, $G_1$ could be varied by changing the substrate's composition and/or polymer concentration. Respectively, $T_\infty$ could be varied by affecting the cell's contractile force generation via myosin phosphorylation, either exogenously by means of drugs or endogenously by the cells as part of a biological process.

\begin{figure}[tb!]
    \centering
    {{\includegraphics[width=\columnwidth]{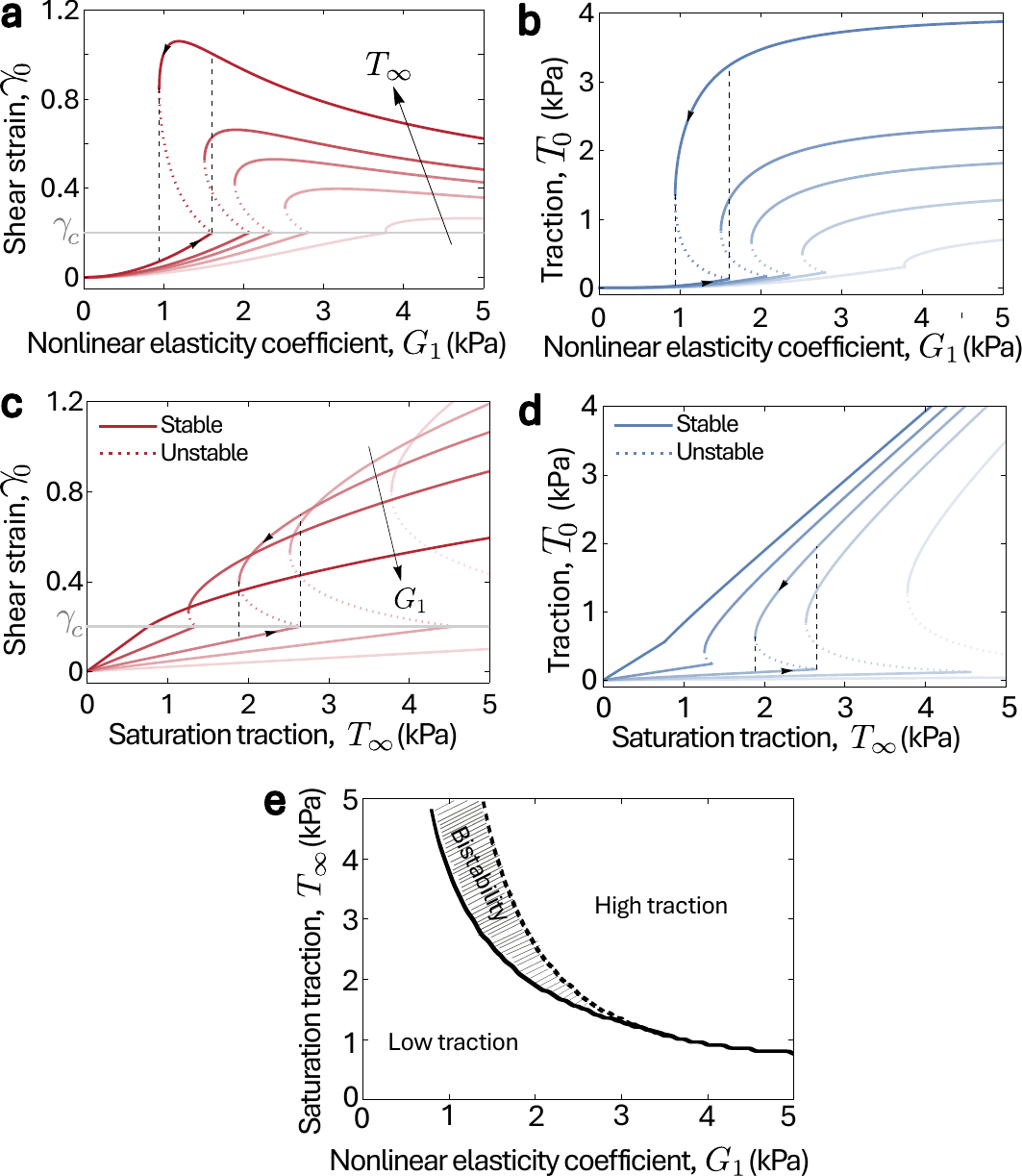}}}
   {\phantomsubcaption\label{Fig strain-G1}}
  {\phantomsubcaption\label{Fig traction-G1}}
  {\phantomsubcaption\label{Fig strain-Tinf}}
  {\phantomsubcaption\label{Fig traction-Tinf}}
  {\phantomsubcaption\label{Fig phasediagrama-G1Tinf}}
    \bfcaption{Traction bistability and hysteresis}{ Substrate strain $\gamma_0$ and traction $T_0$ as a function of the nonlinear elasticity coefficient $G_1$ (\subref*{Fig strain-G1},\subref*{Fig traction-G1}) and the saturation traction $T_\infty$ (\subref*{Fig strain-Tinf},\subref*{Fig traction-Tinf}). In panels \subref*{Fig strain-G1},\subref*{Fig traction-G1}, we vary $T_\infty = 1,1.5,2,2.5,4$ kPa. In panels \subref*{Fig strain-Tinf},\subref*{Fig traction-Tinf}, we vary $G_1 = 1,1.5,2,3,7$ kPa. Solid lines indicate stable states, dotted lines indicate unstable states, and a representative hysteresis cycle is shown with black arrowheads and dashed lines. \subref*{Fig phasediagrama-G1Tinf}, Diagram showing the regions of the low- and high-traction states, as well as their bistable region, as a function of the experimentally tunable parameters $G_1$ and $T_\infty$. Parameter values are $\alpha = 1$, $\beta = 3$, $G^*=1$ kPa, $\gamma_\text{c} = 0.2$, the latter marked by a horizontal gray line in panels \subref*{Fig strain-G1} and \subref*{Fig strain-Tinf}.} 
    \label{fig:All_Hill}
\end{figure}

In \cref{fig:All_Hill}, we plot the stationary solutions combining the linear and nonlinear regimes analyzed above. The strain $\gamma_0$ varies non-monotonically with the nonlinear elasticity coefficient $G_1$ (\cref{Fig strain-G1}). As in the linear regime (\cref{Fig strain-linear}), the non-monotonic behavior is due to the competition between the increase of tractions with substrate stiffness, which leads to greater substrate deformation, and the stiffening of the substrate, which hinders its deformation. The corresponding stationary traction $T_0$, however, increases monotonically with $G_1$ (\cref{Fig traction-G1}).

Interestingly, the linear and nonlinear stable solutions can coexist in a range of $G_1$ values (\cref{Fig strain-G1,Fig traction-G1}). This bistable coexistence of linear and nonlinear solutions happens also as a function of the saturation traction $T_\infty$ (\cref{Fig strain-Tinf,Fig traction-Tinf}). Thus, upon increasing either substrate stiffness $G_1$ or cellular contractility $T_\infty$, cells can discontinuously jump from a low- to a high-traction state (\cref{Fig traction-G1,Fig traction-Tinf}). Moreover, this traction bistability produces a hysteresis loop (black arrowheads and dashed lines in \cref{Fig strain-G1,Fig traction-G1,Fig strain-Tinf,Fig traction-Tinf}). Compared to the transition from low to high tractions, upon reducing $G_1$ or $T_\infty$ starting from the high-traction state, cells will transition to the low-traction state at a lower value of $G_1$ or $T_\infty$. We summarize our findings in \cref{Fig phasediagrama-G1Tinf}, which shows the parameter regions of low and high traction, as well as the bistable region in which both states are possible.

\vspace{0.4cm}
\textbf{Discussion.} By modeling the positive feedback between cellular traction and substrate stiffness, we found a region of bistability between low- and high-traction states. As a result, when varying either the substrate's nonlinear elasticity or cellular contractility, we predict that cells can experience a switch-like discontinuous transition between low and high tractions. This prediction could be tested in experiments that measure cellular tractions on nonlinear substrates \cite{Toyjanova2014,Hall2016,Steinwachs2016,Dong2017,Han2018,Song2020,Bohringer2024,Sarnighausen2025}. In such experiments, the traction transition could be realized either by varying the concentration and/or the crosslinking of the substrate's polymers, or by tuning cellular contractility using drugs that target myosin activity. Below, we discuss possible biological implications of the traction transition, which could apply both to single cells and cell collectives.

The sudden increase in traction that we predict might trigger tissue spreading through an active wetting transition, which takes place above a critical traction \cite{Perez-Gonzalez2019,Alert2020}. The traction transition could thus trigger processes such as embryo implantation, where tractions compete with tissue surface tension to drive spreading \cite{Cavanaugh2025}. In cancer, the stiffening of the extracellular matrix around a tumor might induce the traction transition and thus promote tumor invasion \cite{Mierke2008,Levental2009,Koch2012,Kraning-Rush2012,Krajina2021,Meng2024}. Similarly, matrix deposition around a wound could induce the traction transition, which could initiate the phenotypic transformation of fibroblasts into myofibroblasts \cite{Alisafaei2025,Hong2025} and promote collective cell migration to heal the wound \cite{Friedl2009}. Finally, our findings suggest a physical mechanism whereby the stiffening of the underlying head mesoderm could trigger the migration of neural crest cells during embryo development \cite{Barriga2018}.
Future work could explore whether our results could also explain the emergence of self-generated stiffness gradients \cite{Shellard2021}, where the front of a cell cluster exerts stronger tractions and thus locally stiffnens the substrate.

Our prediction of traction bistability is consistent with recent observations of bursts of high tractions in natural killer cells moving through stiff matrices \cite{Bohringer2024}. Bistability in traction was also recently predicted to emerge from the interaction between focal-adhesion proteins and substrate elasticity \cite{Liu2025}. Our work provides an alternative mechanism for bistability that relies on the nonlinear elastic behavior of the substrate. Interestingly, the bistable behavior might provide robustness to cellular tractions against ECM stiffness variations when cells migrate through heterogeneous environments. For example, bistability ensures that a transition from low to high tractions is not immediately reverted by a small local decrease in ECM stiffness.

More broadly, our work shows how the two-way mechanical interaction between cells and the ECM can strongly regulate cell behavior. Whereas we focused on the interplay between traction forces and substrate elasticity, future works could combine our findings with many other aspects of cell-ECM interactions, such as couplings between ECM strain and cell polarity \cite{Adar2021,Adar2022,Adar2024,Zemel2010}, matrix alignment and cellular contractility \cite{Ahmadzadeh2017,Palmquist2022}, and matrix deposition by cells \cite{Cicconofri2024,Bell2025,Moghe2025}. Other possible extensions include accounting for additional ECM rheological behaviors such as viscoelasticity \cite{Chaudhuri2020,Chaudhuri2015,Bennett2018,Adebowale2021,Krajina2021,Clark2022,Huerta-Lopez2024,Charbonier2025,Courbot2025,Villacrosa-Ribas2026} and plasticity \cite{Vader2009,Baker2015,Kim2017,Ban2018,Krajina2021}. Finally, our work also offers a way to address the role of strain stiffening on cell motility on compliant substrates, including its spontaneous initiation \cite{Etienne2024}, stick-slip motion \cite{Sens2013}, and the biphasic velocity-stiffness relation \cite{Chelly2022}.

\bigskip
\textbf{Acknowledgements.}  This paper was funded by Ministerio de Ciencia, Innovaci\'{o}n y Universidades (MICIU) through a Ph.D. Fellowship FPU19/05492 and EST24/00183 to I.P-J., as well as by a complementary grant from G-Research (February 2024 Grant Winners); by the Generalitat de Catalunya (AGAUR SGR-2017-1061, SGR-2021-00450 and ICREA Acad\`{e}mia awards to J.C.); by the Spanish Ministry for Science and Innovation MICCINN/FEDER (PID2019-108842GB-C21 and PID2022-137713NB-C22 to J.C. and I.P-J.). I.P-J. thanks the Max Planck Institute for the Physics of Complex Systems for hosting a research visit. R.A. acknowledges funding from the European Union through the ERC Starting Grant ``Living\_Fluctuations'' (No. 101114584).

\medskip
\textbf{Author contributions.} R.A. and I.P-J proposed, developed, and solved the model. All authors contributed to developing and interpreting the theory and wrote the paper.


\bibliography{Traction_stiffness_new,Preprints} 

\clearpage
\onecolumngrid
\section*{Supplemental Material}
\renewcommand{\thesection}{\Roman{section}}
\renewcommand{\thesubsection}{\Roman{subsection}}
\renewcommand{\thesubsubsection}{\Roman{subsection}.\arabic{subsubsection}}
\renewcommand{\thetable}{S\Roman{table}} 
\renewcommand{\theHtable}{S\Roman{table}} 
\renewcommand{\theequation}{S\arabic{equation}}
\renewcommand{\thefigure}{S\arabic{figure}}  
\renewcommand{\theHfigure}{S\arabic{figure}} 
\setcounter{secnumdepth}{2}
\setcounter{table}{0}
\setcounter{equation}{0}
\setcounter{figure}{0}
\setcounter{section}{0}
\setcounter{subsection}{0}


Here, we generalize the stability analysis of the nonlinear stationary traction solutions to nonuniform, space-dependent perturbations. The nonuniform traction perturbations produce depth-dependent strains, which we determine in \cref{sec_SI:stability}. In \cref{sec_SI:diffusion}, we include a spatial coupling in the traction relaxation dynamics, which accounts for the tendency of nearby cells to align.

\subsection{Depth-dependent substrate strain}
\label{sec_SI:stability}
\begin{figure}[htb!]
    \centering
    {{\includegraphics[width=0.45\columnwidth]{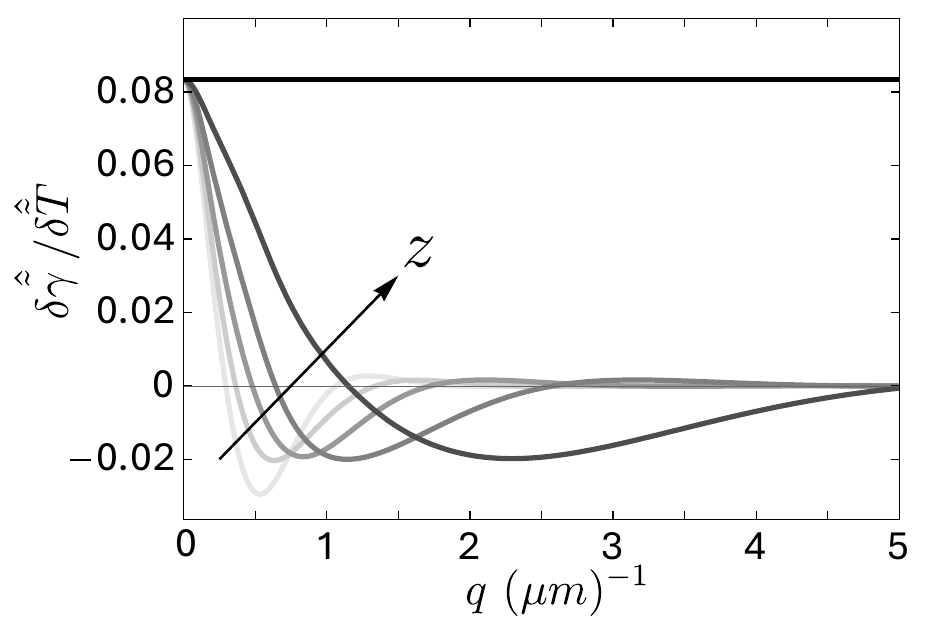}}}
    \bfcaption{Strain perturbation as a function of the wavenumber}{ Traction perturbations of different wavenumber $q$ cause substrate strain perturbations that depend on the vertical coordinate $z$. The plot shows values of $z=0,1,2,3,4,5$ $\mu$m. At the substrate's surface, $z=h = 5$ $\mu$m (black line), the ratio $\delta \tilde{\gamma}_{xz}/\delta\tilde T$ is independent of $q$. Hence, the surface strain has the same spatial profile as the traction $T(x)$ that induces it. Other parameters values are $\alpha=2$, $\delta \tilde{T} = 1$ kPa$\cdot \mu$m$\cdot$h, $K=20$ kPa and $G_1 (\gamma_{xz}^0)^\alpha=2$ kPa.}
    \label{fig:gamma_xz_q_dep}
\end{figure}

Let $\gamma_{xz}^0$ and $T_0$ be the stationary solutions for the shear strain and the traction in the nonlinear regime, with shear modulus $G = G_1\gamma_{xz}^\alpha$ from \cref{eq:G(gamma)}. We then introduce a space-dependent traction perturbation, $T(x) = T_0+\delta T(x)$, which induces depth-dependent substrate strains $\gamma_{xz}(x,z) = \gamma_{xz}^0+\delta \gamma_{xz}(x,z)$. To obtain them, we solve the full system \crefrange{eq:force_balance}{eq:BC}. To this end, we first compute the stress in \cref{eq:stress}, for which we obtain
\begin{align} 
    \sigma_{xx}^0+\delta\sigma_{xx} &= \left(K+\frac{4G_1(\gamma_{xz}^0+\delta\gamma_{xz})^\alpha}{3}\right)(\gamma_{xx}^0+\delta\gamma_{xx}) + \left(K-\frac{2G_1(\gamma_{xz}^0+\delta\gamma_{xz})^\alpha}{3}\right)(\gamma_{zz}^0+\delta\gamma_{zz}), \\ 
    \sigma_{xz}^0+ \delta\sigma_{xz} &= 2G_1(\gamma_{xz}^0+\delta\gamma_{xz})^{\alpha+1} \\
    \sigma_{zz}^0+\delta\sigma_{zz} &= \left(K-\frac{2G_1(\gamma_{xz}^0+\delta\gamma_{xz})^\alpha}{3}\right)(\gamma_{xx}^0+\delta\gamma_{xx}) + \left(K+\frac{4G_1(\gamma_{xz}^0+\delta\gamma_{xz})^\alpha}{3}\right)(\gamma_{zz}^0+\delta\gamma_{zz}). 
\end{align}
To linear order, $(\gamma_{xz}^0+\delta\gamma_{xz})^\alpha \approx (\gamma_{xz}^0)^\alpha + \alpha (\gamma_{xz}^0)^{\alpha-1}\delta \gamma_{xz} $, and since $\gamma_{xx}^0=\gamma_{zz}^0=0$, we have
\begin{align} 
    \delta\sigma_{xx} &\approx C_1\delta\gamma_{xx} + (C_3-C_2)\delta\gamma_{zz}, \\ 
    \delta\sigma_{xz}&\approx 2C_2\delta\gamma_{xz}, \\
    \delta\sigma_{zz} &\approx (C_3-C_2)\delta\gamma_{xx} + C_1\delta\gamma_{zz},
\end{align}
where we defined the constants $C_1 = K + \frac{4}{3}G_1(\gamma_{xz}^0)^\alpha$, $C_2 = G_1(\alpha + 1)(\gamma_{xz}^0)^\alpha$ and $C_3 = K + G_1(\gamma_{xz}^0)^\alpha (\alpha + 1/3)$, so that $C_3-C_2 =  K-\frac{2}{3}G_1(\gamma_{xz}^0)^\alpha$. Thus, the linearized system of equations for the perturbations of the stress and their boundary conditions, from \crefrange{eq:force_balance}{eq:BC}, read
\begin{align} 
    &C_1 \partial_x (\delta \gamma_{xx}) + (C_3-C_2) \partial_x (\delta \gamma_{zz}) + 2C_2 \partial_z (\delta \gamma_{xz}) \approx 0, \\
    &2C_2 \partial_x (\delta \gamma_{xz}) + (C_3-C_2) \partial_z (\delta \gamma_{xx}) + C_1 \partial_z (\delta \gamma_{zz}) \approx 0,  \\ 
    &u_x\big|_{z=0} \approx 0, \quad u_x\big|_{z=0} \approx 0, \\    
    & \left[ (C_3-C_2) \delta\gamma_{xx} + C_1\delta\gamma_{zz} \right] \big|_{z=h} \approx 0, \quad 2 C_2 \delta \gamma_{xz}\big|_{z=h} \approx \delta T. 
\end{align}
Introducing the definition of the strain in terms of the displacements, $\delta \gamma_{xx} =\partial_x (\delta u_x)$, $\delta \gamma_{xz} =\frac{1}{2}(\partial_x (\delta u_z)+\partial_z (\delta u_x))$ and $\delta \gamma_{zz} =\partial_z (\delta u_z)$, we obtain
\begin{align}  
    &C_1  \partial_x^2 (\delta u_x) + (C_3-C_2) \partial_x\partial_z (\delta u_z) + C_2(\partial_z \partial_x (\delta u_z) + \partial_z^2 (\delta u_x)) \approx 0,  \\
    &C_2 (\partial_x^2(\delta u_z) + \partial_x\partial_z (\delta u_x)) + (C_3-C_2) \partial_z\partial_x (\delta u_x) + C_1 \partial_z^2 (\delta u_z) \approx 0, \\ 
    &u_x\big|_{z=0} \approx 0, \quad u_x\big|_{z=0} \approx 0,  \\    
    &\left[ (C_3-C_2) \partial_x (\delta u_x) + C_1\partial_z (\delta u_z) \right] \big|_{z=h} \approx 0,  \quad C_2 \left[ \partial_x (\delta u_z) + \partial_z (\delta u_x) \right] \big|_{z=h} \approx \delta T.
\end{align}
Transforming to Fourier components along the $x$ coordinate with wavenumber $q$, indicated by a tilde, these equations read
\begin{align} 
    &-C_1 q^2 \delta\tilde{u}_x + C_2 \partial_z^2 (\delta\tilde{u}_x) + iqC_3 \partial_z (\delta \tilde{u}_z) \approx 0,  \\ 
    &C_1 \partial_z^2 (\delta \tilde{u}_z) -C_2 q^2 \delta \tilde{u}_z + iqC_3 \partial_z (\delta\tilde{u}_x) \approx 0, \\ 
    &\delta\tilde{u}_x\big|_{z=0} \approx 0,\quad  \delta \tilde{u}_z\big|_{z=0} \approx 0,  \\
    &\left[ (C_3 - C_2)i q\delta\tilde{u}_x  + C_1 \partial_z (\delta \tilde{u}_z) \right] \big|_{z=h} \approx 0, \quad
    C_2 \left[iq \delta \tilde{u}_z + \partial_z (\delta\tilde{u}_x) \right] \big|_{z=h} \approx \delta \tilde{T}. \label{eq:systemFourier}
\end{align}
The solution for this system is a combination of exponentials, $\delta\tilde{u}_x  \approx \sum_{i=1}^4 A_i e^{\lambda_i z} $ and $\delta \tilde{u}_z  \approx \sum_{i=1}^4\mu_i A_i e^{\lambda_i z}$, with $\lambda_i$ and $\mu_i$ functions of $C_1,C_2,C_3$ and $q$ given by
\begin{align} 
   \lambda_1 &= -\lambda_2 = -f_1(C_1,C_2,C_3)q, \\ 
   \lambda_3 &= -\lambda_4= -f_2(C_1,C_2,C_3)q, \\
   \mu_1 &= -\mu_2 = i f_1(C_1,C_2,C_3)g_1(C_1,C_2,C_3) \equiv i h_1(C_1,C_2,C_3), \\ 
   \mu_3 &= -\mu_4= i f_2(C_1,C_2,C_3)g_2(C_1,C_2,C_3) \equiv i h_2(C_1,C_2,C_3), 
\end{align}
where $h_i \equiv f_i g_i$ (with $i=1,2$) and 
\begin{align} 
   f_1 &= \sqrt{\frac{C_1^2+C_2^2-C_3^2 - \sqrt{k}}{2C_1C_2}},\quad f_2 = \sqrt{\frac{C_1^2+C_2^2-C_3^2 + \sqrt{k}}{2C_1C_2}}, \\
   g_1 &= \frac{C_1^2-C_2^2-C_3^2 + \sqrt{k}}{2C_2C_3}, \quad g_2 = \frac{C_1^2-C_2^2-C_3^2 - \sqrt{k}}{2C_2C_3}, 
\end{align}
with $k = (C_1-C_2-C_3)(C_1+C_2-C_3)(C_1-C_2+C_3)(C_1+C_2+C_3)$. Substituting $C_i$ with the corresponding expressions in terms of the bulk and the shear modulus $K$ and $G$, respectively, we obtain $k = -\frac{16}{9} G_1^2 (\gamma_{xz}^0)^{2\alpha} (G_1(\gamma_{xz}^0)^\alpha + 3 K) \alpha [3 K + G_1(\gamma_{xz}^0)^\alpha(4 + 3 \alpha)]$, which implies that $k$ is negative given that $G_1, K,\alpha,\gamma_{xz}^0>0$. Therefore, the functions $f_1$ and $f_2$, which are related to the length-scale in the exponentials, are imaginary. Writing explicitly $\delta\tilde{u}_x$ and $\delta\tilde{u}_x$, one obtains 
\begin{align} 
   \delta\tilde{u}_x  &\approx A_{1} e^{-f_1 q z} + A_{2} e^{f_1 q z} + A_{3} e^{-f_2 q z} + A_{4} e^{f_2 q z} \nonumber \\ &= (-A_{1} + A_{2}) \sinh{(f_1 q z)} + (A_{1} + A_{2}) \cosh{(f_1 q z)} + (-A_{3} + A_{4}) \sinh{(f_2 q z)} +(A_{3} + A_{4}) \cosh{(f_2 q z)}  \nonumber \\ &\equiv B_1 \sinh{(f_1 q z)} + B_2 \cosh{(f_1 q z)} + B_3 \sinh{(f_2 q z)} + B_4 \cosh{(f_2 q z)},  \\
   \delta \tilde{u}_z  &\approx i h_1 ( A_{1} e^{-f_1 q z} - A_{2} e^{f_1 q z} ) + i h_2 (A_{3} e^{-f_2 q z} - A_{4} e^{f_2 q z} ) \nonumber \\ &= i h_1 (-A_{1} - A_{2}) \sinh{(f_1 q z)} + i h_1 (A_{1} - A_{2}) \cosh{(f_1 q z)} + i h_2 (-A_{3} - A_{4}) \sinh{(f_2 q z)} + i h_2 (A_{3} - A_{4}) \cosh{(f_2 q z)}  \nonumber \\ &\equiv -i \Big[  h_1 \big( B_2 \sinh{(f_1 q z)} +  B_1 \cosh{(f_1 q z)} \big) +  h_2 \big( B_4 \sinh{(f_2 q z)} +  B_3 \cosh{(f_2 q z)} \big) \Big], 
\end{align}
where $B_{i}(C_1,C_2,C_3,q,\delta \tilde{T},h)$ are determined by the boundary conditions of \cref{eq:systemFourier}. We see that $\delta\tilde{u}_x$ is real and $\delta \tilde{u}_z$ is imaginary. Hence, the strain perturbation $\delta \tilde{\gamma}_{xz} = \frac{1}{2}[ \partial_z (\delta\tilde{u}_x) + i q\delta \tilde{u}_z]$ is real and it depends on the depth coordinate $z$ (\cref{fig:gamma_xz_q_dep}).

\subsection{Adding a spatial coupling of tractions}
\label{sec_SI:diffusion}

We now extend the traction dynamics \cref{eq:dynamics} to spatially non-uniform states, introducing a simple diffusive coupling, to account, for instance, for the alignment tendency between the polarity of neighboring cells:
\begin{equation} \label{eq:spatial_coupling}
    \partial_t T = -\frac{1}{\tau} \left[ T-T_\text{target} \left(G \big|_{z=h} \right) - D\nabla^2 T \right].
\end{equation}
This form of the spatial coupling follows from modeling a cell layer as a an active polar fluid, in which active tractions are proportional to the tissue polarity field, $\mathbf{T}=T_\text{a} \mathbf{p}$ \cite{Perez-Gonzalez2019,Alert2020}. In this theory, polarity gradients are penalized according to the Frank elastic energy of polar liquid crystals \cite{Perez-Gonzalez2019,Alert2020}, which gives rise to the spatial coupling in \cref{eq:spatial_coupling}.

\begin{figure}[htb!]
    \centering
    {{\includegraphics[width=0.7\columnwidth]{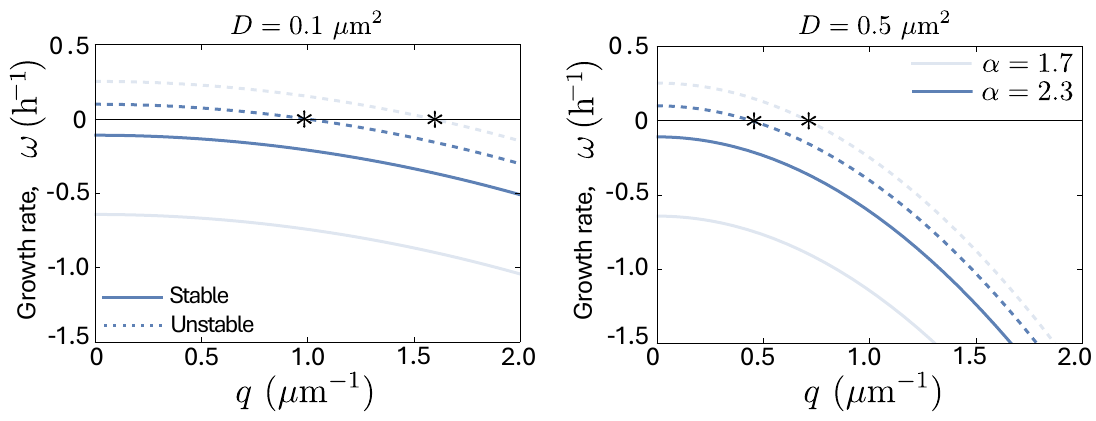}}}
    {\phantomsubcaption\label{Fig dispersionD01}}
    {\phantomsubcaption\label{Fig dispersionD05}}
    \bfcaption{Growth rate of traction perturbations with spatial coupling}{ Growth rate $\omega(q)$ for diffusion coefficients $D=0.1$ $\mu$m$^2$ (\subref*{Fig dispersionD01}) and $D=0.5$ $\mu$m$^2$ (\subref*{Fig dispersionD05}), where solid and dashed lines correspond to the stable (high-traction) and the unstable (low-traction) solutions in \cref{Fig equation-nonlinear-alpha}, respectively, for $\alpha = 1.7$ (light blue) and $\alpha = 2.3$ (dark blue). Thus, values of $\omega(D=0)$ correspond to the $\gamma_0$ solutions in \cref{Fig equation-nonlinear-alpha} (filled and empty dots). For $\alpha = 1.7$, the stable solution is $\gamma_0 = 0.225$ ($\omega(D=0)=-0.643$) and the unstable one $\gamma_0 = 0.038$ ($\omega(D=0) = 0.252$). In this last case, $q^* = 1.588\;\mu\text{m}^{-1}$ (\subref*{Fig dispersionD01}) and $q^* = 0.710\;\mu\text{m}^{-1}$ (\subref*{Fig dispersionD05}), marked by star symbols in both plots. For $\alpha = 2.3$, the stable solution is $\gamma_0 = 0.240$ ($\omega(D=0) = -0.110$) and the unstable one $\gamma_0 = 0.204$ ($\omega(D=0) = 0.099$). In this last case, $q^* = 0.993 \;\mu\text{m}^{-1}$ (\subref*{Fig dispersionD01}) and $q^* =0.444\;\mu\text{m}^{-1}$ (\subref*{Fig dispersionD05}), also  marked by star symbols in both plots.}
    \label{fig:dispersion}
\end{figure}

In the presence of such spatial coupling, the growth rate of traction perturbations around the nonlinear stationary states becomes
\begin{align} \label{eq:growth_rate_spatial_coupling}
    \omega(q) = \omega(D=0) - \frac{D}{\tau}q^2,
\end{align}
where $\omega(D=0)$ is given in \cref{eq:growth-rate}. \cref{eq:growth_rate_spatial_coupling} predicts that, for the low-traction solution (see \cref{Fig equation-nonlinear-alpha,Fig equation-nonlinear-beta}), traction perturbations with wavenumber $q<q^* \equiv  \sqrt{\omega(D=0)\tau/D}$ are unstable (\cref{fig:dispersion}). Thus, the system will transition to the high-traction state at length scales $\lambda\gtrsim \lambda^*\equiv 1/q^*$. Taking ranges of parameter values from previous work \cite{Perez-Gonzalez2019}, we estimate $2.8\;\mu\text{m} < \lambda^* < 9.7\;\mu\text{m}$. Thus, we expect unstable perturbations to take place at length scales longer than a few micrometers, and therefore the low-traction to high-traction transition should be visible at tissue scales. Moreover, we estimate the value of the growth rate to be $\omega^{-1}\sim 2$ h, and hence we expect that the transition would occur over a time scale of a few hours.

\end{document}